\documentclass[aps,twocolumn,prl,10pt,superscriptaddress,floatfix,final]{revtex4-1}
\usepackage{graphicx, bm, amsmath, amsfonts, amssymb, color}
\usepackage[%
  breaklinks=true,
  colorlinks=true,
  linkcolor=blue,
  urlcolor=blue,
  citecolor=blue
]{hyperref}

\newcommand{\Change}[1]{\color{black}#1\color{black}}

\begin{document}

\title{Probing two-level systems with electron spin inversion recovery of defects at the Si/SiO$_2$ interface}
\author{M. Belli}
\affiliation{CNR-IMM, Unit of Agrate Brianza, Via C. Olivetti, 2, 20864, Agrate Brianza (MB), Italy}
\author{M. Fanciulli}
\email{marco.fanciulli@unimib.it}
\affiliation{Universit\`{a} degli Studi di Milano-Bicocca, Dipartmento di Scienza dei Materiali, Via R. Cozzi 53, 20126, Milano, Italy}
\affiliation{CNR-IMM, Unit of Agrate Brianza, Via C. Olivetti, 2, 20864, Agrate Brianza (MB), Italy}
\author{R. de Sousa}
\affiliation{Department of Physics and Astronomy, University of Victoria, Victoria, British Columbia, Canada V8W 2Y2}
\date{\today}

\begin{abstract}
The main feature of amorphous materials is the presence of excess vibrational modes at low energies, giving rise to the so called ``boson peak'' in neutron and optical spectroscopy. 
These same modes manifest themselves as two level systems (TLSs) causing noise and decoherence in qubits and other sensitive devices. Here we present an experiment that uses the spin relaxation of dangling bonds at the Si/(amorphous)SiO$_2$ interface as a probe of TLSs. We introduce a model that is able to explain the observed non-exponential electron spin inversion recovery and provides a measure of the degree of spatial localization and concentration of the TLSs close to the interface, their maximum energy and its temperature dependence.
\end{abstract}
\maketitle

Several properties of amorphous materials can be explained by assuming the presence of additional low energy vibrational modes on top of the usual phonon density of states. In neutron scattering and Raman spectroscopy these modes appear as a universal boson peak with average energy increasing with temperature \cite{Grigera2003, Shintani2008}. At low temperatures, these modes give rise to an anomalous contribution to the specific heat. A convenient assumption is to model the excess modes at the low energy tail of the boson peak as an ensemble of tunnelling two-level systems (TLSs), each with energy splitting $E$. Assuming their energy density scales as a power law with exponent $\alpha$ [$\rho(E)\propto E^\alpha$] leads to specific heat scaling as $T^{1+\alpha}$ \cite{Anderson1972, Phillips1987}. The coefficient $\alpha$ gives a measure of the degree of amorphousness of the material. 

The TLSs are often responsible for the origin of noise, decoherence, and dielectric energy loss in all kinds of devices for solid state quantum computation, including superconducting Josephson devices  \cite{Martinis2005, Gordon2014} and spin qubits \cite{Desousa2007}.  As these devices are generally made from high quality materials, the TLSs usually appear close to surfaces and interfaces, where the degree of crystallinity is quite hard to control. 
\Change{Significant progress has been achieved with the use of superconducting resonators for TLS spectroscopy in the microwave range \cite{Shalibo2010, Grabovskij2012, Skacel2015, Sarabi2016}. These experiments measured the TLS energy-area density in a large number of junctions, thin films, surfaces and interfaces made of amorphous silicon and aluminum oxide. In all cases the TLS density appeared to be close to $\rho(E/h=5~{\rm GHz})\approx 0.1/[(\mu{\rm m})^{2}h\times{\rm GHz}]$ \cite{Grabovskij2012, Skacel2015}.
However, similar to measurements of the Boson peak at thin films \cite{Steurer2008, Zink2006}, the experiments with superconducting resonators lack the energy and temperature bandwidth to measure properties such as the exponent $\alpha$  and the temperature dependence of $\rho(E)$.} 

Here we describe an experiment that uses dangling-bond spins as a probe of TLSs at the Si/SiO$_2$ interface. Unsaturated dangling bonds (DBs), generically called
P$_{\text{b}}$-centers, appear at the Si/SiO$_2$ interfaces \cite{Nishi1971, Caplan1979, Pointdexter1981}. Their structure is quite well understood \cite{Stirling2000}.
We measure spin-lattice relaxation of the DB spin magnetization, $\langle S_z(t)\rangle$, using inversion-recovery experiments with echo detection.  We show that $\langle S_z(t)\rangle$ approaches thermal equilibrium in a highly non-exponential fashion, leading to a wealth of information on the spatial distribution and energetics of TLSs nearby the DB spin. The signal intensity is measurable thanks to a nanostructuring of the interface into nanowires, instead of a flat surface, greatly increasing the surface-to-volume ratio \cite{Fanciulli16}.

It is in fact well known that DBs can act as a probe of TLSs because their spin relaxation rate $1/T_1$ is strongly dominated by TLS dynamics, even at higher temperatures \cite{Stutzmann1983, Askew1986}. However, previous experiments \cite{Askew1986} were unable to interpret the long time non-exponential decay. Below we describe our experiment and propose a  theoretical model based on a Poisson distribution of TLSs within a radius of each DB. This model is able to capture the long time non-exponential dynamics thus allowing the extraction of much more information on TLS parameters than previous approaches. As a result we are able to obtain a clear picture of TLSs at the interface, including  
the measurement of their degree of spatial localization, one of the most important unsolved problems in the physics of the boson peak. 

\begin{figure}
\includegraphics[width=0.48\textwidth]{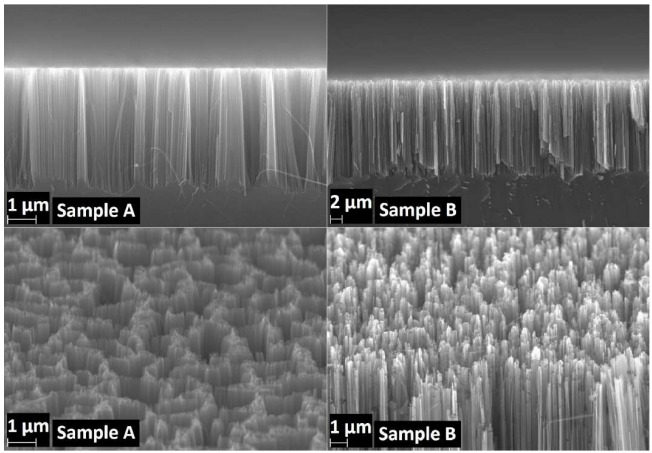}
\caption{\label{Figure1} Scanning Electron Microscope images of the two systems under investigation. Images on the left refer to sample A, while images on the right refer to sample B. The images were taken on twin samples obtained from the same batches of the ones used for magnetic resonance investigations.}
\end{figure}

\emph{Experiment.--} Silicon nanowires were prepared by a metal-assisted chemical etching (MACE) process starting from two different types of seed metal deposited on intrinsic $[$001$]$ silicon. Details of the two samples under investigation are reported in Table \ref{tab1}. For sample A, pinholes in a $3.8$ nm thick gold layer were used to realize the nanowires. For sample B, the seed metal consisted in Ag nanoparticles, deposited by electroless deposition, as explained in Ref.\ \onlinecite{Fanciulli16}, where also details of the etching process are reported. \Change{At the end of the MACE process, the metal was removed from the Si nanowires. The efficiency of the metal removal has been investigated by energy Dispersive X-ray Spectrometry (EDS). 
Within the sensitivity limitation of EDS, of the order of 1000 ppm in weight, SiNWs produced with the Ag NPs (sample B) were found free of metallic particles, while those produced with Au (sample A) revealed a contamination only on the tips of the nanowires, therefore not affecting most of the SiNWs surface. } The full process led to a dense carpet of straight silicon nanowires, fully passivated with H and with structural parameters dependent on the process details (Fig.~\ref{Figure1}). 
The two samples were chosen out of many different batches for the present investigation. Both samples were annealed in vacuum for $15$ minutes at $550^{\circ}$C to induce depassivation of surface defects from the naturally hydrogen-passivated state formed during the MACE process. Details of the depassivation process have been reported elsewhere \cite{Fanciulli16}.

The samples were then characterized by continuous-wave and pulsed
electron paramagnetic resonance (Bruker Elexsys E580 system, X-band). After the depassivation treatment the typical electron paramagnetic resonance signal of DB defects at the interface, the so called P$_b$ defects, was observed to increase \cite{Fanciulli16}. 
Inversion recovery experiments, with the magnetic field H$\parallel [001]$, were performed in the temperature range 4-300 K to obtain information on the relaxation rate. A typical example of an inversion recovery curve is reported in Figure 2, together with a fit attempt with a single exponential recovery.

\begin{table}
\caption{Structural characteristics of the investigated SiNW samples and average distance and concentration of the corresponding P$_{\text{b}}$-like defects. $S/V$ is an estimate of the surface-to-volume increase factor with respect to the case of the flat surface of the bulk.}
\label{tab1}
\begin{tabular}{|c|c|c|} \hline
            & Sample A                                & Sample B                          \\ \hline
Catalyzer   & Au layer                         & Ag NPs                                   \\
Length      & $(4.2 \pm 0.3)$ $\mu$m           & $(17 \pm 1)$ $\mu$m                      \\
Diam. range & $8$ nm - $30$ nm                 & $50$ nm - $200$ nm                       \\
  $<r>$     & $3.7(1)$ nm                      & $3.89(5)$ nm                             \\
  $\left[P_{\text{b}}\right]$  & $8.0(8)\times 10^{11}$ cm$^{-2}$ & $7.2(2)\times 10^{11}$ cm$^{-2}$ \\
  $S/V$       &  $\sim 800$                      & $\sim 3000$                              \\ \hline 
\end{tabular}
\end{table}

 Such a model evidently fails, especially at low temperature, though the resulting thermal trend of the spin-lattice relaxation rate determined assuming a single exponential recovery, may allow comparisons with data reported in the literature. Generally, the inefficiency of a single exponential recovery fit is neglected and the analysis focuses on the thermal variation of the resulting spin-lattice relaxation rate $1/T_{1}$, which is reported to follow a power-law trend $\propto T^{2+\alpha}$ with $\alpha$ in the range $\sim 0.3 - 1.5$ \cite{Askew1986}. The lowest value reported to the authors' knowledge is $\alpha = -0.2$, in Ref.\ \onlinecite{Stutzmann1983}. In our case, the fit would result in even lower $\alpha$ values ranging from $-0.45$ to $-0.55$, which are quite unusual, at least for bulk materials. Fitting attempts with a stretched exponential recovery model were attempted and seemed indeed more successful, though they essentially shift the whole information on the dominant relaxation mechanisms into the temperature dependence of other two physical quantities: the stretched relaxation rate and the stretched exponent, which require further interpretation. We think that a deeper understanding of the non-exponential recovery is necessary. The temperature dependence of the spin lattice relaxation rate, obtained either by a single exponential or by a stretched inversion recovery, was attributed to the presence of TLSs.  We need therefore a broader theoretical framework modelling the role of the TLSs already at the level of the recovery curves.

\begin{figure}
\includegraphics[width=0.48\textwidth]{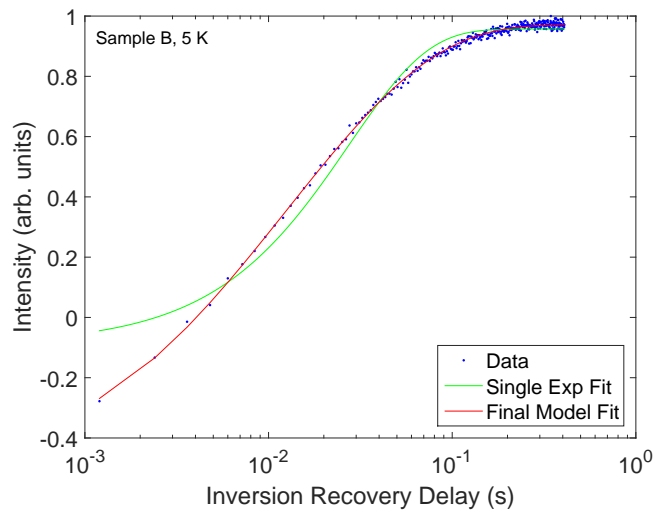}
\caption{\label{Figure2} (Color online) Comparison between a single exponential inversion recovery fit at 5 K for sample B and a fit according to the model outlined in Eq.~(\ref{eq:IRmodel}). \Change{Fits for sample A had slightly larger $\chi_{r}^{2}$ (Table~\ref{tab2}).}}
\end{figure}

\emph{Theoretical model.--} Previous models for DB spin relaxation \cite{Askew1986, Desousa2007} assumed a \emph{low disorder prescription}:  Each DB spin was assumed to relax through cross-relaxation with \emph{\Change{exactly } one TLS}.  \Change{If the TLSs are randomly distributed at the interface, the number of TLSs within a ``coupling radius'' of the P$_b$ center will follow a Poisson distribution,
\begin{equation}
p_n=\frac{\bar{N}^n}{n!}\textrm{e}^{-\bar{N}}.
\label{p_n}
\end{equation}
Here $p_n$ denotes the fraction of P$_b$ centers with $n$ TLSs coupled to it, and $\bar{N}$ the average number of TLSs coupled to each P$_b$ DB. 
It should be emphasized that such a model implies a much greater level of disorder than previous models that assumed a Kronecker delta distribution $p_n=\delta_{n,1}$ \cite{Desousa2007, Askew1986}. We note that the Poisson model is quite different than this previous model even when $\bar{N}=1$, since in this case a large fraction of the P$_b$ centers $p_0=\textrm{e}^{-1}=37$\% are not relaxing at all, while $p_2+p_3\cdots=26$\% relax much faster due to coupling to two or more TLSs.}

Here we propose a model for this \emph{high disorder prescription} and show that it yields much better fits for the non-exponential time decay of P$_b$ spins. Assume the TLSs do not interact with each other and are independently distributed with energy splitting in the interval $E\in [0, E_{{\rm max}}]$, each with density proportional to $E^\alpha$. The average spin magnetization for a DB interacting with $n$ TLSs is given by
\begin{eqnarray}
\langle S_z(t)\rangle_n &=& \frac{\int_{0}^{E_{{\rm max}}}dE_1 E_{1}^{\alpha} \cdots \int_{0}^{E_{{\rm max}}}dE_n E_{n}^{\alpha} \textrm{e}^{-\sum_{i=1}^{n}\Gamma(E_i,T)t}
}{\int_{0}^{E_{{\rm max}}}dE_1 E_{1}^{\alpha} \cdots \int_{0}^{E_{{\rm max}}}dE_n E_{n}^{\alpha}}\nonumber\\
&=&\left(\langle S_z(t)\rangle_1\right)^{n},
\end{eqnarray}
where $\Gamma(E_i,T)$ is the spin relaxation rate for a DB interacting with one TLS of energy $E_i$, and 
\begin{equation}
\langle S_z(t)\rangle_1=\frac{\alpha+1}{E_{{\rm max}}^{\alpha+1}} \int_{0}^{E_{{\rm max}}}dE E^{\alpha} \textrm{e}^{-\Gamma(E,T)t}
\label{sz1}
\end{equation}
the associated magnetization decay. 
Taking an average with $p_n$ as in Eq.~(\ref{p_n}) leads to
\begin{equation}
\langle\langle S_z(t)\rangle\rangle =  \sum_{n=0}^{\infty} p_n \langle S_z(t)\rangle_n = \exp{\left\{-\bar{N}\left[1-\langle S_z(t)\rangle_1\right]\right\}}.
\label{szdavg}
\end{equation}
This expression shows that the Poisson distribution of TLSs makes DB spin relaxation highly non-exponential in time.

In order to complete the model we need to obtain a suitable expression for the relaxation rate $\Gamma(E,T)$, the rate for a DB spin to achieve thermal equilibrium with a single TLS of energy $E$. The mechanism is based on spin-orbit induced cross-relaxation \cite{Desousa2007}. When a TLS switches, the local environment around the DB spin fluctuates; spin-orbit coupling translates this switch into a fluctuating magnetic field that flips the spin. 

The energy eigenstates of each TLS are denoted $\mid\pm\rangle$ with energies $\pm E/2$, and the transition rate for a TLS to switch from state $\mid\pm\rangle$ to state $\mid\mp\rangle$ is denoted $r_\pm$ (the subscript refers to the initial state in the transition). 
When the magnetic field is low so that DB Zeeman energy is much less than both $k_BT$ and $E$, the rate for cross-relaxation is well approximated by \cite{Desousa2007}  
\begin{equation}
\Gamma_{\pm\uparrow}=\Gamma_{\pm\downarrow}\approx A^{2}r_{\pm},
\label{gar}
\end{equation}
with $A\ll 1$ a dimensionless spin-orbit coupling parameter. Here $\Gamma_{+\uparrow}$ denotes the rate for a \emph{cross} switch from the TLS-DB state $|+,\uparrow\rangle$ into the state  $|-,\downarrow\rangle$. These cross rates are much stronger than non-cross spin flips because they couple states that are \emph{not the time reversal of each other} ($|+,\uparrow\rangle$ is not the time reversal of $|-,\downarrow\rangle$). \Change{Only when the DB Zeeman energy is quite large other non-cross spin-flip mechanisms become important. Magnetic fields in the Tesla range are required to polarize the DB spins so that their magnetic noise due to coupling to TLSs is suppressed.}

The thermalized rate $\Gamma(E,T)$ is obtained by averaging over TLS and DB spin states with their corresponding Boltzmann occupations. 
Denote $p(i|\sigma)$ the probability of finding TLS in state $i=+,-$ given that the DB spin is known to be in state $\sigma=\uparrow,\downarrow$. 
In the limit of DB Zeeman energy much smaller than $k_BT,E$ we get $p(i|\uparrow)\approx p(i|\downarrow)$ hence
\begin{eqnarray}
\Gamma(E,T)&=&\sum_{i=+,-}\sum_{\sigma=\uparrow,\downarrow} p(i|\sigma)\Gamma_{i\sigma}\approx 2A^2\left(p_{+}r_{+}+p_{-}r_{-}\right)\nonumber\\
&=&4A^2\frac{r_{+}r_{-}}{r_{+}+r_{-}}.
\label{gammabar}
\end{eqnarray}

Note how the DB spin relaxation rate $\Gamma(E,T)$ is solely determined by the TLS rates $r_{\pm}$. To describe the physics up to quite high temperatures we generalized the theory for $r_{\pm}$ described in \cite{Desousa2007} to processes involving one and two acoustic phonons. The final result is 
\begin{eqnarray}
\Gamma(E,T)&=&a \left\{ \frac{E/k_B}{\sinh{(\beta E)}}+b \frac{(E/k_B)^5}{1+\textrm{e}^{\beta E}}\left[0.00714+\frac{2930}{(\beta E)^7}
\right.\right.\nonumber\\
&&\left.\left.\times\left(1+\frac{\beta E}{2}+\frac{(\beta E)^2}{10}+\frac{(\beta E)^3}{100}\right)\right]\right\},
\label{Gamma}
\end{eqnarray}
where $\beta=1/(k_B T)$, and $a$ and $b$ model the linear and quadratic dependence of TLS parameters on the phonon dilation strain, respectively. Therefore, $a$
models the efficacy of TLS flipping following the emission/absorption
of a single acoustic phonon and $b$ models the same effect involving two phonons.
Equation~(\ref{Gamma}) was plugged into Eqs.~(\ref{sz1})~and~(\ref{szdavg}) leading to an explicit analytic expression for the measured inversion recovery curve,
\begin{widetext}
\begin{equation}
\frac{I}{I_0} = 1-2 \exp{\left\{-\bar{N}\left[ 1-\left( \left( \beta E_{{\rm max}}\right) ^{-(\alpha +1)}\right) \left( \displaystyle\int\limits_0^{\left( \beta E_{{\rm max}} \right) ^{\alpha +1}} dx\ \mathrm{e}^{-\Gamma \left( x^{\frac{1}{\alpha +1}},T\right) t} \right) \right]\right\}}. \label{eq:IRmodel}
\end{equation}
\end{widetext}

We stress again the highly non-exponential form of the model. This expression has five fitting parameters: $a,b, \alpha, \bar{N}, E_{{\rm max}}$. The fitting was done by assuming the first four parameters independent of temperature, with $E_{{\rm max}}$ temperature dependent according to
\begin{equation}
E_{{\rm max}}(T)=c + d T + f T^2 + g T^3.
\label{Emax}
\end{equation}
The assumed polynomial fit for the $T$ dependence of $E_{{\rm max}}$ may be seen as an approximation to the complex TLS thermal activation. 

The fit results are reported in Table~\ref{tab2}. \Change{We note that the fit parameters $a$, $b$, and $\alpha$ come out quite different for the two samples. Given that these parameters are highly sensitive to the TLS morphology we conclude that the TLS structure is quite different for samples A and B. This is not surprising given the drastic difference in sample preparation method. }
Differences can be observed also in the parameters describing the thermal evolution of $E_{{\rm max}}$, though the two leading terms, $c$ and $d$, are relatively similar. This implies that the difference between the two samples is more relevant in the higher temperature range.

\begin{table}
\caption{Fitted parameters according to model described in Equations \ref{Gamma}, \ref{eq:IRmodel}, \ref{Emax}.}
\label{tab2}
\begin{tabular}{|c|c|c|}  \hline            
                   & Sample A                    & Sample B                \\ \hline
$a$                & $2337$($3$) Hz/K   & $244.2$($2$) Hz/K          \\
$b/k_\mathrm{B}^4$ & ($58000 \pm 8000$) J$^{-4}$ & ($1800 \pm 300$) J$^{-4}$                 \\
$c$                & $613$($1$) $\mu$eV & $409$($2$) $\mu$eV             \\
$d$                & $569.5$($1$) $\mu$eV/K      & $558.9$($1$) $\mu$eV/K                \\
$f$                & $-619$($2$) neV/K$^2$       & $-2035$($1$) neV/K$^2$   \\
$g$                & $0.263$($9$) neV/K$^3$      & $1.530$($6$) neV/K$^3$                \\
$\alpha$           & $3.074$($6$)                & $2.127$($5$)                           \\
$\bar{N}$                & $0.7290$($1$)               & $0.68900$($9$)         \\ \hline 
$\chi^2_r$         & $1.3157$        & $1.2003$                               \\ \hline 
\end{tabular}
\end{table}

\Change{\emph{Estimate of TLS spatial extent.--} 
The P$_b$ center acts as a ``local probe'' that senses TLSs in its immediate neighborhood. As a result, our fits do not yield direct information on the total area density for TLSs, which is in principle unrelated to our measurements of the P$_b$ density shown in Table~\ref{tab1}.}

\Change{To interpret our results further, we assume the TLS energy density at $E_0 = h\times 5~{\rm GHz}$ is equal to the one estimated for amorphous SiO films \cite{Skacel2015}, $\rho(E_0)= 0.1/[(\mu{\rm m})^{2}h\times{\rm GHz}]$, and that this value can be extrapolated to other energies using our measurements of $\alpha$. This leads to the following total TLS area density,
\begin{equation}
    \sigma_{TLS}=\int_{0}^{E_{{\rm max}}}dE \rho(E_0)\left(\frac{E}{E_0}\right)^{\alpha}. 
\end{equation}
Using our fit parameters at $T=5~K$ we get 
$\sigma_{TLS}\approx 2\times 10^{16}$~cm$^{-2}$ for sample A and $\sigma_{TLS}\approx 6\times 10^{13}$~cm$^{-2}$ for sample B. This shows that sample A has significantly larger disorder at the surface of its nanowires.}

\Change{Since it is well known that the P$_b$ center is highly localized (within a few Angstroms \cite{Pointdexter1981}) the interaction range for its coupling to TLSs can be interpreted as being equal to the spatial extension of the TLS itself, $l_{TLS}$. The average number of TLSs coupling to the P$_b$ can then be written as $\bar{N}\approx (\sigma_{TLS}/t_I)l_{TLS}^{3}$, with $t_I\approx 40$~\AA~the interface thickness \cite{Pan}. Using the fit values $\bar{N}\approx 0.7$ we get 
$l_{TLS}\approx (\bar{N}t_I/\sigma_{TLS})^{1/3}= 3$~\AA~ for sample A and $l_{TLS}\approx 17$~\AA~ for sample B. } This shows that the TLSs at the interface are similar to dangling-bonds defects in that they are localized within a few atomic sites.

\emph{Conclusions.--} In conclusion, we exploited the high interface area of silicon nanowires to detect, with good signal-to-noise ratio, the electron spin inversion recovery of P$_b$ centers at the Si/SiO$_2$ interface. 
A novel model was developed to describe the non-exponential character of the inversion recovery.
\Change{Fitting the data with this model demonstrated that dangling-bonds can act as a ``local probe'' for the properties of amorphous TLSs.} 

\Change{The proposed method, when combined with measurements of TLS energy density in the microwave range, lead to the conclusion that TLSs at the interface are localized atomic vibrations extending over a few lattice sites ($3-17$~\AA). These studies can be  extended to other relevant systems. } 
A comparison of the results with theoretical models of specific TLSs, \Change{such as bistable atomic distortions associated to vacancies or other point defects in the oxide, } may lead to the still missing identification of the microscopic nature of the TLSs.
\begin{acknowledgments}
M.F. and M.B. acknowledge financial support from the CARIPLO Foundation (Italy), ELIOS project, and the Italian Ministry of Defense, QUDEC project. 
R.d.S. acknowledges financial support from NSERC (Canada) through its Discovery (RGPIN-2015- 03938) and Collaborative Research and Development programs (CRDPJ 478366-14). We thank J. Fabian for useful discussions. 
\end{acknowledgments}

\bibliography{references}

\end{document}